\documentclass{article}
\newcommand{\be}{\begin{equation}}
\newcommand{\ee}{\end{equation}}

\newcommand{\ndt}{\noindent}

\def\bea{\begin{eqnarray}}
\def\eea{\end{eqnarray}}
\def\beas{\begin{eqnarray*}}
\def\eeas{\end{eqnarray*}}
\def\sla{\raise.15ex\hbox{$/$}\kern-.57em}

\newcommand{\del}{{\partial}}

\newcommand\fr[1]{\frac{1}{#1}}
\newcommand{\nn}{\nonumber}
\newcommand{\orderof}{\ensuremath{\mathcal{O}}}

\usepackage{amsmath}
\usepackage{latexsym}
\usepackage{graphicx}
\begin{document}
\begin{titlepage}
\vskip 1cm
\centerline{\LARGE{\bf {Light-cone gravity in dS$_4$}}}

\vskip 1.5cm
\centerline{{Sudarshan Ananth and Mahendra Mali}} 
\vskip 1cm
\centerline{\em  Indian Institute of Science Education and Research}
\centerline{\em Pune 411008}

\vskip 1.5cm

\centerline{\bf {Abstract}}
\vskip .5cm
\ndt We derive a closed form expression for the light-cone Lagrangian describing pure gravity on a four-dimensional de Sitter background. We provide a perturbative expansion of this Lagrangian to cubic order in the fields. 
\vfill

\end{titlepage}

\section{Introduction}
\vskip 0.3cm

Attempts to unite quantum mechanics and the general theory of relativity result in divergences which are difficult to treat. In this context, the resemblance of perturbative gravity to gauge theory is particularly striking. Specifically, the KLT relations~\cite{KLT} equate tree-level scattering amplitudes in pure gravity to the square of tree-level amplitudes in Yang-Mills theory. Over the past decade, our understanding of these relations and their origin has improved greatly. In particular, the Lagrangian origin of this connection is now well established~\cite{AT}. At the lowest interaction order - the cubic vertex - we now possess a plethora of interesting perturbative links between interacting theories of arbitrary spin~\cite{AA} (making the cubic KLT relations merely one in a family). 

\vskip 0.3cm

\ndt Almost all these perturbative ties have been derived on flat spacetime backgrounds so it is natural to ask whether these relations or their equivalents exist in curved spacetime backgrounds. While it is not clear what the stringy origin would be for a Yang-Mills $\sim$ gravity link in curved spacetime, the question itself is interesting and well posed within the framework of quantum field theory.  
\vskip 0.3cm
\ndt This paper is a companion paper to our earlier formulation of pure gravity on an AdS$_4$ background~\cite{ASM}. Although the differences in treatment from the anti-de Sitter case are not significant, we feel the closed form result for the light-cone gravity action in de Sitter will prove extremely useful in studies using perturbative quantum field theory (particularly in the context of cosmology). This closed form and the vertices that result from its expansion are also essential in the investigations of the perturbative ties described above. Thus, in this paper, we formulate pure gravity in light-cone gauge, on a dS$_4$ background. 

\vskip 0.3cm
\ndt On a tangential note, the surprising ultra-violet behavior of $\mathcal N=8$ supergravity is thought to stem from the better-than-expected behavior of pure gravity in the ultra-violet regime~\cite{BD}.  This motivates our study of pure gravity on various backgrounds, its nature and the detailed structure of its Lagrangian in terms of the physical degrees of freedom. 

\vskip 0.3cm
\ndt Our approach in this paper is similar in spirit to~\cite{ASM,BCL}. We make suitable gauge choices and eliminate the unphysical degrees of freedom using the light-cone constraint relations. This will result in a closed form expression for the action of pure gravity in a de Sitter background. We also provide a perturbative expansion of this action to first order in the gravitational coupling constant (the cubic interaction vertex).

\vskip 0.3cm

\section{Einstein gravity}
\label{sec:eingra}
\vskip 0.3cm

\ndt The Einstein-Hilbert action, describing pure gravity, reads
\bea
\label{eh}
S_{EH}=\int\,{d^4}x\,L\,=\,\frac{1}{2\,\kappa^2}\,\int\,{d^4}x\,{\sqrt {-g}}\,(\,{\mathcal R}\, - 2\,\Lambda\,)\ ,
\eea
\ndt where $g=\det{g_{\mu\nu}}$, $\mathcal R$ is the scalar curvature, $\Lambda$ the cosmological constant of dS$_4$ and $\kappa^2=8\pi G$, the gravitational coupling constant.

\vskip 0.3cm

\ndt This theory has been studied previously, in light-cone gauge, on both AdS$_4$ and flat backgrounds~\cite{ASM,BCL}. In this paper, we formulate pure gravity in $dS_4$ characterized by a cosmological constant, $\Lambda$. This will involve changes from both the flat spactime and anti-de Sitter approaches referred to above and we comment on these departures as and when they occur.
 
\vskip 0.5cm

\subsection{ de Sitter Space }

\ndt de Sitter spacetime is a maximally symmetric Lorentzian space with positive (constant) curvature~\cite{WDS}. It is a solution of the equation
\bea
R_{\mu\nu}-\frac{1}{2}g_{\mu\nu}+\Lambda g_{\mu\nu} =0\ ,
\eea
with cosmological constant $\Lambda$. de Sitter space is a hyperboloid embedded in a five dimensional Minkowski space.
\vskip 0.3cm
\ndt Consider a five-dimensional Minkowski spacetime with metric $\eta_{_{MN}}\equiv(-1,1,1,1,1)$ and co-ordinates $\xi^{0}\,,\xi^{1}\,,\xi^{2}\,,\xi^{3}\,, \xi^{4}$. The invariant interval reads  
\bea 
ds^2=-(d\xi^{0})^2+(d\xi^{1})^2+(d\xi^{2})^2+(d\xi^{3})^2+(d\xi^{4})^2\ ,
\eea
with $\xi^{M}\in (-\infty,+\infty), M=0\ldots4 $. de Sitter space is the hypersurface 
\bea \label{dsm}
-(\xi^{0})^2+(\xi^{1})^2+(\xi^{2})^2+(\xi^{3})^2+(\xi^{4})^2=l^2=H^{-2}\ ,
\eea
where we have related the radius of de Sitter space $l$ to the Hubble constant $H=l^{-1}$. A solution of $(\ref{dsm})$ is
\bea 
-(H\,\xi^0)^2\,+\,(H\,\xi^4)^2&=& 1\,-\, (H\,x^i)^2\, e^{2Ht}\ , \nn \\
(H\,\xi^1)^2\,+\,(H\,\xi^2)^2\,+\,(H\,\xi^3)^2&=&(H\,x^i)^2\, e^{2Ht}\ ,
\eea 
\ndt with 
\bea \label {coordinates}
H\,\xi^0&=& \sinh{(H\,t)}+\frac{{(H\,x^i)}^2}{2}e^{Ht}\ , \nn\\
H\,\xi^i&=& H\,x^i\, e^{H\,t}\ ,  \nn\\
H\,\xi^4&=& -\, \cosh{(H\,t)}+\frac{{(H\,x^i)}^2}{2}e^{Ht}\ ,
\eea
where  $x^i \, \in \, (-\infty,+\infty), i=1,2,3\,$ and $\, t \, \in \, (-\infty,+\infty).$\
\ndt In terms of these new coordinates, the induced metric is 
\bea 
\label{eppmetric}
ds^2=-(dt)^2+e^{2Ht}\{(dx^1)^2+(dx^2)^2+(dx^3)^2\}\ .
\eea
Our choice of coordinates in $(\ref{coordinates})$ impose the following constraint
\bea
 -\, \xi^0\,+\,\xi^4\,=\,-\,\frac{1}{H}\,e^{Ht}\,\leq\,0\, \implies \, \xi^0\,\geq\,\xi^4 \ ,
\eea
implying that we are only covering one half of the de Sitter space (expanding Poincar\'{e} patch of dS). Within this patch, we define conformal time by
\bea
H\eta=e^{-Ht}\ ,
\eea
which modifies the metric $(\ref{eppmetric})$ to 
\bea
\label{met}
ds^2=\frac{1}{H^2\eta^2}(-d\eta^2+(dx^1)^2+(dx^2)^2+(dx^3)^2)\ .
\eea
Note that the conformal time runs from $\eta\,=\,+\infty$ ($t\,=\,-\infty$) to $\eta\,=\,0$ ($t\,=\,+\infty$). We work in this expanding Poincar\'{e} patch of dS$_4$ but we could equally well have worked with the other patch (contracting patch of dS). 

\vskip 0.3cm

\section{Light-cone formulation of pure gravity on dS$_4$}
We start with the  metric of $(\ref{met})$  which reads
\bea 
g^{(0)}_{\mu\nu}=\frac{1}{H^2\eta^2}\eta_{\mu\nu}\  ,
\eea
where $\eta_{\mu\nu \,}=(-1,1,1,1,)$ is the four-dimensional Minkowski metric. We now introduce light-cone coordinates, $x^\mu \equiv (x^+,x^-,x^i)$  where\\
\bea
x^{\pm}=\frac{\eta \pm x^{3}}{\sqrt{2}}\ ,
\eea
and $i=1,2$ label the transverse directions. The coordinate $x^+$ is now the evolution parameter. In terms of these coordinates, the Minkwoski metric is $\eta^{L.C.}_{\mu\nu}$ (which is off-diagonal for the $+$, $-$ coordinates and diagonal for the $i$ directions). We also define
\bea
X=x^++x^-\ .
\eea
Our metric now reads 
\bea 
g^{(0)}_{\mu\nu}=\frac{2}{H^2 X^2}\eta^{L.C.}_{\mu\nu}\ .
\eea
The cosmological constant of $dS_4$ is 
\bea
\label{cc}
\Lambda=3H^2\ .
\eea

\vskip 0.3cm

\subsection{Light-cone action}

\vskip 0.1cm

\ndt We now proceed to gauge fix the Einstein-Hilbert action and derive a closed form expression for the action in terms of the physical degrees of freedom in the theory. We start with the light-cone gauge choices
\bea
\label{lcg}
g_{--}\,=\,g_{-i}\,=\,0\quad ,\; i=1,2\ .
\eea
Note that these choices are consistent with $g_{\mu\nu}^{(0)}$ since $\eta_{--}=\eta_{-i}=0$. The fourth (and final) gauge choice will be made shortly. The other components of the metric are parametrized as follows.
\bea
\label{gc}
\begin{split}
g_{+-}\,&=\,-\,\frac{2}{H^2 X^2}e^\phi\ , \\
g_{i\,j}\,&=\,\frac{2}{H^2 X^2}e^\psi\,\gamma_{ij}\ .
\end{split}
\eea
\noindent $\phi\,,\,\psi$ are real and $\gamma_{ij}$ is a real, symmetric and unimodular matrix describing the two physical degrees of freedom in the theory. 

\vskip 0.3cm

\noindent In light-cone gauge, a subset of the equations of motion ${\mathcal R}_{\mu\nu}-\fr{2}g_{\mu\nu}\,{\mathcal R}\,=-\Lambda\,g_{\mu\nu}$ represent contraint relations which may be solved. The key difference in dealing with constraint relations in dS$_4$, as opposed to both AdS$_4$ and flat space~\cite{ASM, BCL}  stems from the fact that $X$ in (\ref {gc}) depends on $\partial_-$. Since constraint relations always contain $\partial_-$ we will need integrating factors to solve them. A listing of some useful formulae used in the following is presented in Appendix A. The first constraint relation reads ${\mathcal R}_{--}=0$ and combined with (\ref {gc}) implies that
\bea
\label{con2}
\partial_-\,\phi\,\partial_-\,\psi\,-\,{\partial_-}^2\,\psi\,-\,\frac{1}{2}\,{(\partial_-\,\psi)}^2\,-\,\frac{2}{X}\partial_-\,\phi\,+\,\frac{1}{4}\,\partial_-\,\gamma^{kl}\,\partial_-\,\gamma_{kl}\,=\,0\ 
\eea
This constraint is exactly solvable if we make the following (fourth) gauge choice 
\bea
\label{gc4}
\phi\,=\,\frac{1}{2}\,\psi\ .
\eea
Now $(\ref{con2})$ simplifies to
\bea
\frac{1}{4}\,\partial_-\,\gamma^{kl}\,\partial_-\,\gamma_{kl}\,-\,{\partial_-}^2\,\psi\,-\,\frac{2}{X}\partial_-\,\phi\,=\,0\ ,
\eea
which when multiplied by an integrating factor ($X$) results in
\bea 
\label{psi}
\psi\,=\,\frac{1}{4}\,\frac{1}{\partial_-}\,\left[\frac{1}{X}\,\frac{1}{\partial_-}\,(X\partial_-\,\gamma^{kl}\,\partial_-\,\gamma_{kl})\right]\ ,
\eea
where $\frac{1}{\partial_-}$ is defined following the prescription in~$\cite{SM}$. Notice the difference in structure here (\ref {psi}) from the analogous results in AdS$_4$ and flat space~\cite{ASM,BCL}.

\vskip 0.3cm

\ndt We now move to the second constraint: ${\mathcal R}_{-i}=0$. With use of an integrating factor, $\frac{1}{X^2}$, this yields
\bea
g^{-i}\,=&\,H^2X^2\mathrm{e}^{-\,\phi}\,\frac{1}{\partial_-}\bigg[X^2\,\gamma^{ij}\,\mathrm{e}^{\phi\,-\,2\,\psi}\,\frac{1}{\partial_-}\,{\Big \{}\,\frac{1}{X^2}\,\mathrm{e}^{\psi}\,{\Big (}\,\frac{1}{4}\,\partial_-\,\gamma^{kl}\,\partial_j\,\gamma_{kl}\,-\,\frac{1}{2}\,\partial_-\,\partial_j\,\phi \nn\\
&\,-\,\frac{1}{2}\,\partial_-\,\partial_j\,\psi\,+\,\frac{1}{2}\partial_j\phi\,\partial_-\,\psi\,-\,\frac{2}{X}\partial_j\,\phi\,{\Big )}\,+\,\frac{1}{2X^2}\,\partial_l\,{\Big (}\,\mathrm{e}^{\psi}\,\gamma^{kl}\,\partial_-\,\gamma_{jk}\,{\Big )}\,{\Big \}}\,\bigg]\ .
\eea

\vskip 0.3cm

\ndt Having determined these components of the metric, we turn to the action 
\bea
S=\int d^4x\,{\mathcal L}=\frac{1}{2 \kappa^2}\int d^4x \sqrt{-g} \left( 2 g^{+-} R_{+-} +g^{i j} R_{i j} - 2 \Lambda\right),\ 
\eea
\vskip 0.3cm
which written out explicitly reads
\bea
\label{aaction}
S\,&=&\int d^{4}x \frac{1}{H^2X^2}e^{\psi}\left(\frac{24}{X^2}+4\del_{+}\del_{-}\phi - 2\del_+\psi\del_-\psi - \del_{+}\gamma^{ij}\del_{-}\gamma_{ij}\right) \nonumber \\
&&-\frac{1}{H^2X^2}e^{\phi}\gamma^{ij}\left(2\del_{i}\del_{j}\phi +  \del_{i}\phi\del_{j}\phi - 2\del_{i}\phi\del_{j}\psi - \frac{1}{2}\del_{i}\gamma^{kl}\del_{j}\gamma_{kl} +  \del_{i}\gamma^{kl}\del_{k}\gamma_{jl}\right) \nn \\
&&-\frac{4}{H^2X^2} e^{\phi - 2\psi}\gamma^{ij}\frac{1}{\del_{-}}R_{i}\frac{1}{\del_{-}}R_{j} -\,\frac{8}{H^4X^4}\,e^\psi\,e^\phi\,\Lambda\ ,
\eea
 where 
\bea 
R_i&=&\,\frac{1}{X^2}\,\mathrm{e}^{\psi}\,{\Big (}\,\frac{1}{4}\,\partial_-\,\gamma^{kl}\,\partial_i\,\gamma_{kl}\,-\,\frac{1}{2}\,\partial_-\,\partial_i\,\phi \,-\,\frac{1}{2}\,\partial_-\,\partial_i\,\psi\,+\,\frac{1}{2}\partial_i\phi\,\partial_-\,\psi\,-\,\frac{2}{X}\partial_i\,\phi\,{\Big )}\, \nn\\
&&+\,\frac{1}{2X^2}\,\partial_l\,{\Big (}\,\mathrm{e}^{\psi}\,\gamma^{kl}\,\partial_-\,\gamma_{ik}\,{\Big )}.\
\eea
\vskip 0.3cm
While obtaining this expression, boundary terms have been dropped. This closed form of the action $(\ref{aaction})$ is valid in both patches of de Sitter.

\vskip 0.3cm

\subsection{Perturbative expansion}
We now expand the action in $(\ref{aaction})$ to first order in the gravitational coupling constant. We parametrize $\gamma_{ij}$ as follows.
\bea 
\gamma_{ij}\,=\, (e^H)_{ij}\nn\ ,
\eea
with
\bea 
H\,=\,\begin{pmatrix} h_{11} & h_{12} \\ h_{12} & h_{22} \end{pmatrix}\ .
\eea
$h_{22}\,=\,-\,h_{11}$ ensures that this matrix is traceless. $\psi$ in terms of these fluctuations is 
\bea\label{psifi}
\psi\,=\,-\,\frac{1}{4}\,\frac{1}{\partial_-}\left[\,\frac{1}{X}\,\frac{1}{\partial_-}(X\,\partial_-h_{ij}\,\partial_-h_{ij})\right]\,+\, \orderof(h^4)\ ,
\eea
We re-scale the $h$ field according to 
\bea
h\, \rightarrow\, \frac{1}{\sqrt{2}\,\kappa}\, h\ .
\eea
We now present the kinetic and cubic interaction vertices in the action (\ref {aaction}). 
\bea 
S_2\,=\, \int d^4x \,\mathcal{L}_2\ ,
\eea
where 
\bea 
\mathcal{L}_2 &=& \frac{1}{2H^2X^4}\,\frac{1}{\partial_-}\,(X\partial_-h_{ij}\partial_-h_{ij})\,-\, \frac{1}{2H^2X^3}\,\frac{\partial_+}{\partial_-}\,(X\partial_-h_{ij}\partial_-h_{ij})\nn\\
&&+\,\frac{1}{H^2X^2}\, \partial_+h_{ij}\partial_-h_{ij}\,+\,\frac{1}{2H^2X^2}\,\frac{\partial_i\partial_i}{\partial_-}\,\left[\frac{1}{X}\frac{1}{\partial_-}\,(X\partial_-h_{jk}\partial_-h_{jk})\right]\nn\\
&&-\,\frac{1}{2H^2X^2}\,\partial_ih_{jk}\partial_ih_{jk}\,+\,\frac{1}{H^2X^2}\,\partial_ih_{jk}\partial_jh_{ik}\nn\\
&&+\,\frac{3}{H^2X^4}\,\frac{1}{\partial_-}\,\left[\frac{1}{X}\frac{1}{\partial_-}\,(X\partial_-h_{ij}\partial_-h_{ij})\right]\nn\\
&&-\,\frac{1}{H^2X^2}\,\frac{1}{\partial_-}\,\left(\frac{1}{X^2}\,\partial_j\partial_-h_{ij}\right)\,\frac{1}{\partial_-}\,\left(\frac{1}{X^2}\,\partial_k\partial_-h_{ik}\right)\ .
\eea
\vskip 0.3cm
\ndt From (\ref {gc4}) and the last term in (\ref {aaction}), it is obvious that the cosmological constant $\Lambda$ (\ref {cc}) is always accompnied by $\psi$. Thus, given the structure of (\ref {psifi}), $\Lambda$ only contributes to interaction vertices involving an even number of fields.
\bea 
S_3\,=\, \int d^4x \,\frac{1}{\sqrt{2}}\,\mathcal{L}_3\ ,
\eea
with
\bea
\mathcal{L}_3&=& \kappa\, \Biggr\{ \frac{1}{H^2X^2}\,\frac{1}{\partial_-}\,\left(\frac{1}{X^2}\,\partial_-h_{jk}\partial_ih_{jk}\right)\,\frac{1}{\partial_-}\left(\frac{1}{X^2}\,\partial_l\partial_-h_{il}\right)\nn\\
&&-\,\frac{3}{H^2X^2}\,\frac{1}{\partial_-}\left[\frac{1}{X}\,\frac{\partial_i}{\partial_-}\,(X\partial_-h_{jk}\partial_-h_{jk})\right]\,\frac{1}{\partial_-}\left(\frac{1}{X^2}\,\partial_l\partial_-h_{il}\right)\nn\\
&&-\,\frac{1}{H^2X^2}\,\frac{1}{\partial_-}\left(\frac{1}{X}\,\frac{\partial_i}{\partial_-}\left[\frac{1}{X}\,\frac{1}{\partial_-}(X\partial_-h_{jk}\partial_-h_{jk})\right]\right) \frac{1}{\partial_-}\left(\frac{1}{X^2}\,\partial_l\partial_-h_{il}\right)\nn\\
&&+\,\frac{2}{H^2X^2}\,\frac{1}{\partial_-}\left(\frac{1}{X^2}\,\partial_j h_{jk}\partial_-h_{ik}\right) \frac{1}{\partial_-}\left(\frac{1}{X^2}\,\partial_l\partial_-h_{il}\right)\nn\\ 
&&-\, \frac{1}{H^2X^2}\,\frac{1}{\partial_-}\left( \frac{1}{X^2}\,\partial_j\partial_-h^2_{ij}\right)\frac{1}{\partial_-}\left(\frac{1}{X^2}\,\partial_l\partial_-h_{il}\right)\nn\\
&&+\, \frac{1}{H^2X^2}\,h_{ij}\frac{1}{\partial_-}\left(\frac{1}{X^2}\,\partial_k\partial_-h_{ik}\right)\frac{1}{\partial_-}\left(\frac{1}{X^2}\,\partial_l\partial_-h_{il}\right)\Biggr\}\ .
\eea

\begin{center}
* ~ * ~ *
\end{center}
As expected, both the closed form and the perturbative expansions involve fields tangled with the co-ordinates (here $x^+$ and $x^-$) and involve conformal-like factors. From these expressions, it would be interesting to understand how to extract amplitudes and other related structures. Clearly, it would interesting to extend our work to quartic order in the fields and work out how to deal with the time derivatives ($\partial_+$) that begin to appear then (usually handled using a suitable field redefinition). One interesting but perhaps impractial idea would be to identify a general closed form expression for the action describing light-cone gravity which may be tuned to the space-time of our choice: whether flat, AdS or dS. Finally, such a closed form expression could prove fruitful in identifying the origin of the better than expected ultra-violet behavior seen in pure gravity~\cite{BD}.

\newpage
\appendix 
\section{Useful results}

\beas
\Gamma_{++}^+&=&\frac{1}{2}{g^{+-}[2\partial_+g_{+-}-\partial_-g_{++}]}\\ [\baselineskip]
\Gamma_{+-}^+&=&0\\ [\baselineskip]
\Gamma_{--}^+&=&0\\ [\baselineskip]
\Gamma_{i-}^+&=&0\\ [\baselineskip]
\Gamma_{i+}^+&=&\frac{1}{2}g^{+-}[\partial_ig_{+-}-\partial_-g_{i+}]\\ [\baselineskip]
\Gamma_{ij}^+&=&-\frac{1}{2}g^{+-}\partial_-g_{ij}\\ [\baselineskip]
\Gamma_{--}^-&=&g^{+-}\partial_-g_{+-}\\ [\baselineskip]
\Gamma_{+-}^-&=&\frac{1}{2}\{g^{+-}\partial_-g_{++}+g^{-i}[\partial_-g_{i+}-\partial_ig_{+-}]\}\\ [\baselineskip]
\Gamma_{++}^-&=&\frac{1}{2}\{g^{+-}\partial_+g_{++}+g^{--}[2\partial_+g_{+-}-\partial_-g_{++}]\nonumber\\ 
&&+ g^{-i}[2\partial_+g_{i+}-\partial_ig_{++}]\}\\ [\baselineskip]
\Gamma_{+i}^-&=&\frac{1}{2}\{g^{+-}\partial_ig_{++}+g^{--}[\partial_ig_{+-}-\partial_-g_{i+}]\nonumber\\ [\baselineskip]
&&+g^{-j}[\partial_ig_{+j}+\partial_+g_{ij}-\partial_jg_{+i}]\}\\ [\baselineskip]
\Gamma_{-i}^-&=&\frac{1}{2}\{g^{+-}[\partial_ig_{+-}+\partial_-g_{+i}]+g^{-j}\partial_-g_{ij}\}\\ [\baselineskip]
\Gamma_{ij}^-&=&\frac{1}{2}\{g^{+-}[\partial_ig_{+j}+\partial_jg_{+i}-\partial_+g_{ij}]-g^{--}\partial_-g_{ij}\nonumber\\
&&+g^{-k}[\partial_ig_{kj}+\partial_jg_{ik}-\partial_kg_{ij}]\}\\ [\baselineskip]
\Gamma_{jk}^i&=&\frac{1}{2}\{-g^{-i}\partial_-g_{jk}+g^{im}[\partial_jg_{mk}+\partial_kg_{mj}-\partial_mg_{jk}]\}\\ [\baselineskip]
\Gamma_{-j}^i&=&\frac{1}{2}g^{ik}\partial_-g_{kj}
\eeas
\beas
\Gamma_{+-}^i&=&\frac{1}{2}g^{ij}[\partial_-g_{j+}-\partial_jg_{+-}]\\ [\baselineskip]
\Gamma_{+j}^i&=&\frac{1}{2}\{g^{-i}[\partial_jg_{+-}-\partial_-g_{+j}]+g^{ik}[\partial_jg_{+k}+\partial_+g_{kj}-\partial_kg_{+j}]\}\\ [\baselineskip]
\Gamma_{++}^i&=&\frac{1}{2}\{g^{-i}[2\partial_+g_{+-}-\partial_-g_{++}]+g^{ij}[2\partial_+g_{+j}-\partial_jg_{++}]\}\\ [\baselineskip]
\Gamma_{--}^i&=&0\\ [\baselineskip]
\Gamma_{ij}^j&=&\frac{1}{2}\{-g^{-j}\partial_-g_{ij}+g^{jl}[\partial_jg_{li}+\partial_ig_{lj}-\partial_lg_{ij}]\}\\ [\baselineskip]
\eeas

\vskip 0.1cm

Frequently used quantities
\beas
g^{+-}&=&\,-\frac{H^2X^2}{2}e^{-\phi},\nn \\ 
g^{ij}& =& \,\frac{H^2X^2}{2} e^{-\psi}\gamma^{ij}, \nn \\
\gamma^{ij}&=&\,(e^{-H})_{ij} ,\nn \\
g^{\mu\nu}g_{\mu\rho}&=&\, \delta^{\nu}_{\rho}\implies g^{++} = g^{+i} = 0, \nn \\
g_{+i}&=&\,-g_{+-}g_{ij}g^{-j}, \nn \\
g_{++}&=&\,-\frac{4}{H^4X^4}e^{\psi}g^{--}+\frac{2}{H^2X^2}e^{\phi}g^{-i}g_{+i},\nn \\ 
\gamma^{ij}\gamma_{ij}&=&\,2, \nn \\ 
\gamma^{ij}\del_{k}\gamma_{ij}&=&\, \gamma^{ij}\del_{-}\gamma_{ij}\,=\,0, \nn \\
\sqrt{-g}&=&\,\frac{4}{H^4X^4}e^{\psi}e^{\phi} . \ 
\eeas

\end{document}